\documentclass[twoside,twocolumn,9pt]{article}
\usepackage{extsizes}
\usepackage[super,sort&compress,comma]{natbib} 
\usepackage[version=3]{mhchem}
\usepackage[left=1.5cm, right=1.5cm, top=1.785cm, bottom=2.0cm]{geometry}
\usepackage{balance}
\usepackage{mathptmx}
\usepackage{sectsty}
\usepackage{graphicx} 
\usepackage{lastpage}
\usepackage[format=plain,justification=justified,singlelinecheck=false,font={stretch=1.125,small,sf},labelfont=bf,labelsep=space]{caption}
\usepackage{float}
\usepackage{fancyhdr}
\usepackage{fnpos}
\usepackage[english]{babel}
\usepackage[utf8]{inputenc}
\addto{\captionsenglish}{%
  
}
\usepackage{array}
\usepackage{droidsans}
\usepackage{charter}
\usepackage[T1]{fontenc}
\usepackage[usenames,dvipsnames]{xcolor}
\usepackage{setspace}
\usepackage[compact]{titlesec}
\usepackage{hyperref}

\usepackage[version=3]{mhchem} 
\usepackage{bm} 
\usepackage{natbib}
\usepackage{color}
\usepackage{mathrsfs}
\usepackage{comment}

\definecolor{cream}{RGB}{222,217,201}

\begin{document}

\pagestyle{fancy}
\thispagestyle{plain}
\fancypagestyle{plain}{
\renewcommand{\headrulewidth}{0pt}
}

\newcommand{\pH}{\mathrm{pH}}
\newcommand{\pK}{\mathrm{pK_a}}
\def\bnabla{\mbox{\boldmath $\nabla $}}

\makeFNbottom
\makeatletter
\renewcommand\LARGE{\@setfontsize\LARGE{15pt}{17}}
\renewcommand\Large{\@setfontsize\Large{12pt}{14}}
\renewcommand\large{\@setfontsize\large{10pt}{12}}
\renewcommand\footnotesize{\@setfontsize\footnotesize{7pt}{10}}
\makeatother

\renewcommand{\thefootnote}{\fnsymbol{footnote}}
\renewcommand\footnoterule{\vspace*{1pt}%
\color{cream}\hrule width 3.5in height 0.4pt \color{black}\vspace*{5pt}} 
\setcounter{secnumdepth}{5}

\makeatletter 
\renewcommand\@biblabel[1]{#1}            
\renewcommand\@makefntext[1]%
{\noindent\makebox[0pt][r]{\@thefnmark\,}#1}
\makeatother 
\renewcommand{\figurename}{\small{Fig.}~}
\sectionfont{\sffamily\Large}
\subsectionfont{\normalsize}
\subsubsectionfont{\bf}
\setstretch{1.125} 
\setlength{\skip\footins}{0.8cm}
\setlength{\footnotesep}{0.25cm}
\setlength{\jot}{10pt}
\titlespacing*{\section}{0pt}{4pt}{4pt}
\titlespacing*{\subsection}{0pt}{15pt}{1pt}

\fancyfoot{}
\fancyfoot[LO,RE]{\vspace{-7.1pt}\includegraphics[height=9pt]{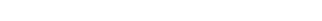}}
\fancyfoot[CO]{\vspace{-7.1pt}\hspace{13.2cm}\includegraphics{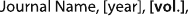}}
\fancyfoot[CE]{\vspace{-7.2pt}\hspace{-14.2cm}\includegraphics{head_foot/RF}}
\fancyfoot[RO]{\footnotesize{\sffamily{1--\pageref{LastPage} ~\textbar  \hspace{2pt}\thepage}}}
\fancyfoot[LE]{\footnotesize{\sffamily{\thepage~\textbar\hspace{3.45cm} 1--\pageref{LastPage}}}}
\fancyhead{}
\renewcommand{\headrulewidth}{0pt} 
\renewcommand{\footrulewidth}{0pt}
\setlength{\arrayrulewidth}{1pt}
\setlength{\columnsep}{6.5mm}
\setlength\bibsep{1pt}

\makeatletter 
\newlength{\figrulesep} 
\setlength{\figrulesep}{0.5\textfloatsep} 

\newcommand{\topfigrule}{\vspace*{-1pt}%
\noindent{\color{cream}\rule[-\figrulesep]{\columnwidth}{1.5pt}} }

\newcommand{\botfigrule}{\vspace*{-2pt}%
\noindent{\color{cream}\rule[\figrulesep]{\columnwidth}{1.5pt}} }

\newcommand{\dblfigrule}{\vspace*{-1pt}%
\noindent{\color{cream}\rule[-\figrulesep]{\textwidth}{1.5pt}} }

\makeatother

\twocolumn[
  \begin{@twocolumnfalse}
{\includegraphics[height=30pt]{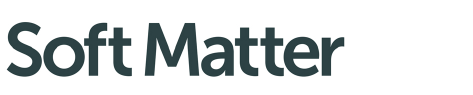}\hfill\raisebox{0pt}[0pt][0pt]{\includegraphics[height=55pt]{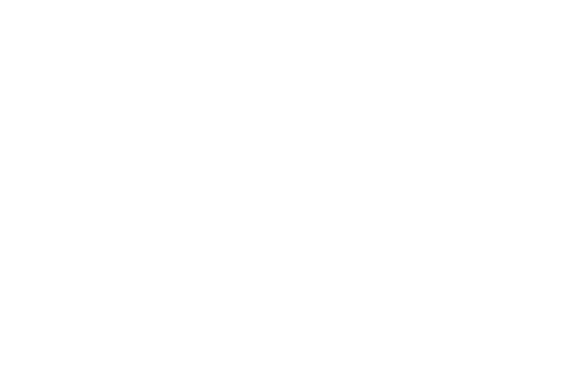}}\\[1ex]
\includegraphics[width=18.5cm]{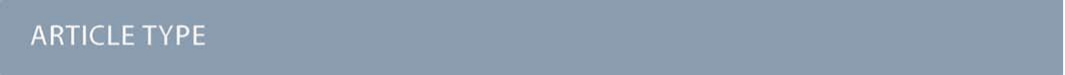}}\par
\vspace{1em}
\sffamily
\begin{tabular}{m{4.5cm} p{13.5cm} }

\includegraphics{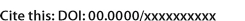} & \noindent\LARGE{\textbf{Electrostatic interaction between SARS-CoV-2 virus and charged electret fibre$^\dag$}} \\
\vspace{0.3cm} & \vspace{0.3cm} \\

 & \noindent\large{Leili Javidpour,\textit{$^{a}$} An\v ze Bo\v zi\v c,\textit{$^{b}$} Ali Naji,\textit{$^{a,c}$} and Rudolf Podgornik$^{\ast}$\textit{$^{d,e,\ddag}$}} \\

\includegraphics{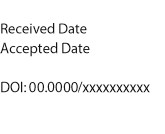} & \noindent\normalsize{While almost any kind of face mask offers some protection against particles and pathogens of different sizes, the most efficient ones make use of a layered structure where one or more layers are electrically charged. These electret layers are essential to efficient filtration of difficult-to-capture small particles, yet the exact nature of electrostatic capture with respect to both the charge on the particles and the electret fibres as well as the effect of immediate environment remains unclear. Here, we explore in detail the electrostatic interaction between the surface of a single charged electret fibre and a model of SARS-CoV-2 virus. Using Poisson-Boltzmann electrostatics coupled to a detailed spike protein charge regulation model, we show how pH and salt concentration drastically change both the scale and the sign of the interaction. Furthermore, the configuration of the few spike proteins closest to the electret fibre turns out to be as important for the strength of the interaction as their total number on the virus envelope, a direct consequence of spike protein charge regulation. The results of our work elucidate the details of virus electrostatics and contribute to the general understanding of efficient virus filtration mechanisms.} \\

\end{tabular}

 \end{@twocolumnfalse} \vspace{0.6cm}

  ]

\renewcommand*\rmdefault{bch}\normalfont\upshape
\rmfamily
\section*{}
\vspace{-1cm}


\footnotetext{\textit{$^{a}$~School of Physics, Institute for Research in Fundamental Sciences (IPM), Tehran, 19395-5531, Iran. }}
\footnotetext{\textit{$^{b}$~Department of Theoretical Physics, Jo\v{z}ef Stefan Institute, SI-1000 Ljubljana, Slovenia. }}
\footnotetext{\textit{$^{c}$~School of Nano Science, Institute for Research in Fundamental Sciences (IPM), Tehran, 19395-5531, Iran. }}
\footnotetext{\textit{$^{d}$~School of Physical Sciences and Kavli Institute for Theoretical Sciences, University of Chinese Academy of Sciences, Beijing 100049, China and Institute of Physics, Chinese Academy of Sciences, Beijing 100190, China; E-mail: podgornikrudolf@ucas.ac.cn}}
\footnotetext{\textit{$^{e}$~Wenzhou Institute of the University of Chinese Academy of Sciences, Wenzhou, Zhejiang 325000, China. }}

\footnotetext{\dag~Electronic Supplementary Information (ESI) available. See DOI: 10.1039/cXsm00000x/}

\footnotetext{\ddag~Also affiliated with  Department of Physics, Faculty of Mathematics and Physics, University of Ljubljana, SI-1000 Ljubljana, Slovenia and Department of Theoretical Physics, Jo\v zef Stefan Institute, SI-1000 Ljubljana, Slovenia.}



\section*{Introduction}

Aerosol spread of respiratory droplets is one of the major pathways for transmission of respiratory diseases~\cite{Kutter2018,Fernstrom2013,laRosa2013,Bozic2020} and an important factor in the recent COVID-19 pandemic caused by the SARS-CoV-2 virus~\cite{Dancer2020,Prather2020,Morawska2020a}. Wearing a face mask importantly limits the spread of droplets and thus significantly reduces transmission~\cite{Prather2020,Morawska2020a,Chu2020} and can lead to milder disease manifestation even in the case of an infection~\cite{Gandhi2020}. While almost any kind of mask aids in reducing the chance of transmission~\cite{Davies2013,Fischer2020,Long2020}, N95 filtering facepiece respirators are typically recommended as personal protective equipment for healthcare professionals due to their high filtering efficiency~\cite{Chu2020,Long2020}. A crucial element of N95 respirators is a filtration layer containing non-woven electrically charged melt-blown polypropylene fibres, which significantly increases the efficiency of filtration, especially in the micrometer range of otherwise difficult-to-capture aerosol particles~\cite{Liao2020,Dyrud1989}. Furthermore, while static charge degradation in the electret layer is a known factor reducing the filtration efficiency~\cite{Liao2020}, recent work shows that it is possible to recharge the masks post-decontamination and recover the filtration efficiency~\cite{Hossain2020}.

Particle filtration is a multiscale process that spans scales from tens-of-micrometers-large respiratory droplets interacting with the filtration layer and then all the way to the nanoscale size of a single pathogen in a bathing solution interacting with a single fibre. While a fibrous filter captures particles efficiently by inertial impaction, interception, and convective Brownian diffusion even in the absence of electrostatic forces, it is the latter---as in the case of electrically charged melt-blown polypropylene fibres---that can significantly increase the efficiency of collection~\cite{Wang2001}. Experimentally, the influence of charged electret filters on collection efficiency is well-studied~\cite{Lundgren1965,Fjeld1988,Romay1998,Ardkapan2014,Chen1998,Tsai2002,Walsh1997}. However, there exist only a few detailed theoretical models describing the mechanisms behind the collection efficiency in electrically charged fibrous layers~\cite{Walsh1997,Kraemer1955, Lathrache1987,Wang2001,Thakur2013}, and none of them address the role particle properties---most importantly its charge---play in this process. These, however, should not be neglected: There are wide differences in charge carried by, for instance, viruses from different species~\cite{Bozic2012,Bozic2017b}, and their electrostatic properties can be further regulated by environmental variables, most notably pH and salt concentration~\cite{Bozic2017,Bozic2018,Nap_Anze_2014}, as well as charge regulation by local electrostatic potential~\cite{Markovich2015,GeneralCR2019,Nap_Anze_2014,Krishnan2017,Atalay2014}. In addition, the nature of the charging process itself makes determination of the net charge on a single fibre difficult, with even the sign of the charge likely varying between individual fibres~\cite{Brown1982,Nifuku2001}, making this yet another variable in an already complex system.

Our work elucidates the electrostatic interactions during the last, nanoscale stage of the filtration process, which has thus far received scant attention. Using a model of a SARS-CoV-2 virus as a case study, we examine the range and the magnitude of its electrostatic interaction with the charged surface of a polypropylene fibre. Bearing in mind the many unknowns in the value and sign of the electret charge and in the details of the composition of the bathing solution, we explore a range of parameters wide enough to capture the most salient features of reality. Our results show that charge regulation, virus geometry close to the fibre, and the environmental parameters are all important factors in determining both the sign and magnitude of the electrostatic interaction, and that manipulation of electrostatic properties in such systems could be used in improving the design of electret filters.

\section*{Theory and Methods}

\subsection*{Virus model}

\begin{figure*}[!htp]
\centering
\includegraphics[width=0.74\linewidth]{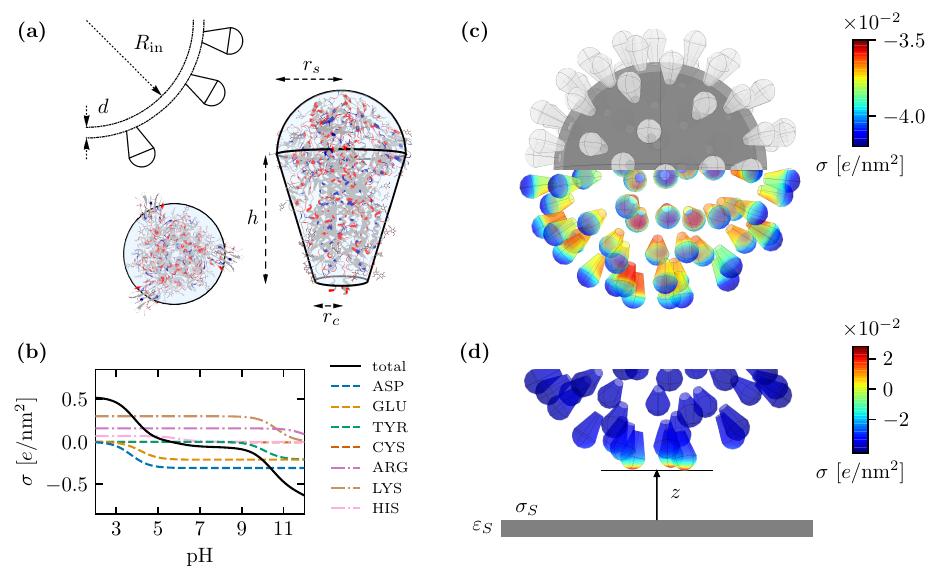}
\caption{Geometry of the virus model and the ice-cream cone model of the SARS-CoV-2 S protein. {\bf (a)} The lipid envelope on which the S proteins are positioned is modelled as two concentric shells with inner radius $R_{in}=40$ nm and thickness $d=4$ nm. The cone model parameters are $r_s=5$~nm, $r_c=2$~nm and $h=10$~nm. Side and top views of the cone also show the experimentally resolved structure of the S protein (PDB:6VXX~\cite{Walls2020}). Red and blue colours denote the acidic and basic amino acid residues, respectively; all other amino acids are shown in gray. Both the S proteins and the lipid membrane have a dielectric permittivity of $\varepsilon=4$. {\bf (b)} Surface charge density on the S protein in absence of local electrostatic effects as a function of pH and the contribution of different acidic and basic amino acids to it. {\bf (c)} Geometry of the virus model (top half of the panel) and charge-regulated surface charge density on the S proteins (bottom half), the only part of the virus carrying charge in our model. Salt concentration of the surrounding medium is $n_0=2$~mM, and $\pH=7$. {\bf (d)} Detail of the virus from panel (c) approaching the electret surface with dielectric constant $\varepsilon_S=2$ and surface charge density $\sigma_S=-1$ $e/\mathrm{nm}^2$, showing the influence of the electret on the surface charge density of the S proteins.
}
\label{fig:1}
\end{figure*}

The SARS-CoV-2 virus consists of a nucleocapsid complex enveloped by an approximately spherical lipid membrane with a number of S proteins arranged on its surface~\cite{Ke2020,Yao2020,Bar2020,Beniac2006}. Since much remains unclear about both the properties of the nucleocapsid as well as the structure of the lipid envelope, and since the major part of interaction of the virus with its surroundings proceeds through the S proteins, we assume the same holds true for its electrostatic interaction. We thus neglect the structure of the virus interior and position the S proteins on a spherical lipid bilayer with inner radius $R_{in}=40$~nm, outer radius $R_{out}=44$~nm~\cite{Yao2020}, and zero surface charge density on both bilayer surfaces (Fig.~\ref{fig:1}a). The lipid envelope is considered a dielectric with $\varepsilon_m=4$, a slightly higher value than was recently measured for DPPC~\cite{Gramse2013}. As the detailed composition of the lipid envelope, which includes membrane proteins and possibly also charged lipids, is unknown, it is only the order of magnitude of $\varepsilon_m$ that matters. We assume that there are no salt ions inside the lipid envelope, but that the envelope is completely permeable to them.

On the outer surface of the lipid envelope we position $N$ S proteins. Given the variability in the number of S proteins in coronaviruses~\cite{Bar2020,Beniac2006,Ke2020,Yao2020}, we vary $N$ between $60$ and $100$, although it has recently been shown that this number can be as low as $20$~\cite{Ke2020,Yao2020}. We model the shape and charge of the S proteins in their closed form based on the Protein Data Bank entry PDB:6VXX~\cite{Walls2020}. The shape of the S protein is modelled as an ``ice-cream cone''---a truncated cone capped with a hemisphere (Fig~\ref{fig:1}a). For the base radius (attached to the surface of the lipid envelope), we use $r_c=2$~nm, and the height of the cone is $h=10$~nm. The top radius of the cone---and simultaneously the radius of the top hemisphere---is $r_s=5$~nm. This gives for the total surface area of the model protein $A_{S}=399.24$~nm$^2$. The S proteins are treated as impermeable to salt ions with a dielectric constant $\varepsilon_{S}=4$.

Unlike simple viruses, whose proteins form capsids with icosahedral symmetry~\cite{Siber2020,Zandi2020}, S proteins of coronaviruses do not possess any apparent spatial order---at least not one with high symmetry~\cite{Bar2020,Ke2020,Yao2020}. A likely reason is that the S protein location within the envelope is constrained by interactions with and between membrane proteins~\cite{Neuman2006,Neuman2011}. For this reason, we consider different possible configurations of S proteins, generated using Mitchell algorithm~\cite{Mitchell1991} (an example is shown in Fig.~\ref{fig:1}c). This preserves (some) minimum distance between the proteins but does not impose any particular spatial order on them~\cite{Bozic2019}. Due to some degree of randomness inherent in such distributions, we use between $18$ and $25$ different distributions for each set of parameter values studied, sufficient to yield reasonably small errors in the calculated force (shown on an example in Fig.~S2, ESI$\dag$). Recent work also shows that the orientation of the S proteins within the membrane is flexible and they do not need to point radially outward~\cite{Ke2020,Yao2020}, while experiments on other types of viruses reveal their flexibility and deformation on approach to a substrate~\cite{Jimenez2018}. Nevertheless, in our model we assume a rigid structure of the S protein layer in order to reduce the amount of system parameters. The flexibility of S proteins could become particularly important upon their contact with the electret, which is, however, outside the present scope of our study.

\subsection*{Charge regulation of spike proteins}

The S proteins of SARS-CoV-2 virus carry a number of amino acid (AA) residues---the unique components among each of the $20$ AAs---which can protonate or deprotonate, acquiring positive or negative charge, respectively. To describe this process, we use Ninham-Parsegian charge regulation theory that connects the local electrostatic potential $\varphi(\mathbf{r})$ with the surface charge density $\sigma(\mathbf{r})$ on the S proteins~\cite{GeneralCR2019,Nap_Anze_2014}. The effective charge of an AA at position ${\bf r}$ is~\cite{Bozic2017,Bozic2018}
\begin{equation}
\label{eq:pn}
q_{\pm}({\bf r})=\frac{\pm e}{1+\exp(\pm\ln10(\pH-\pK)\pm\beta e\varphi(\mathbf{r}))},
\end{equation}
where $\pK$ is its chemical dissociation constant and $\pH$ takes on the bulk value of the electrolyte solution. Equation~\eqref{eq:pn} is obviously just a modified Henderson-Hasselbalch equation with a local electrostatic $\pK$ shift of $\pK\rightarrow\pK\mp(\beta e\varphi/\ln10)$~\cite{Krishnan2017}.

To determine the charge on the S proteins, we extract solvent-accessible ionizable AA residues from the protein structural data (ESI$\dag$). In the absence of a detailed atomic resolution of the AA charging process~\cite{Adhikari2020}, we consider a coarse-grained model where the charge contributed by the AAs is assumed to be homogeneously smeared over the surface of the model S protein and we attribute to each AA its bulk $\mathrm{pK_a}$ value (Table~SI, ESI$\dag$). The total charge-regulated surface charge density on the S proteins is thus a sum of the contributions of the solvent-accessible ionizable AAs:
\begin{eqnarray}
\label{eq:sigma}
\sigma(r)&=&\!\!\!\!\!\!\!\!\!\sum_{i= {\rm ASP},{\rm GLU}, {\rm TYR}, {\rm CYS}}\frac{q_{-}/A_{S}}{1+e^{-\beta e \varphi(r)-\ln10\,(\pH-\mathrm{pK_{a}^{(i)}})}}  \nonumber\\
&+&\!\!\!\sum_{i={\rm ARG}, {\rm LYS}, {\rm HIS}}\frac{q_{+}/A_{S}}{1+e^{\beta e \varphi(r)+\ln10\,(\pH-\mathrm{pK_{a}^{(i)}})}}.
\end{eqnarray}
Here, $\pm$ pertains to deprotonated (ASP, GLU, TYR, and CYS) and protonated (ARG, LYS, and HIS) AAs, respectively, and $r$ runs over the surface of the protein. This approach is similar to the one used previously by Nap et al.~\cite{Nap_Anze_2014}.

\subsection*{Electret model}

Since the diameter of a polypropylene fibre is on the order of $1$ to $10$~$\mu$m~\cite{Siag1994,Yim2020}, up to two orders of magnitude larger than the average diameter of SARS-CoV-2, we model a single electret fibre as a planar wall and assume that at close distance, any geometrical effects of the fibre shape can be neglected. The electret has a dielectric constant of $\varepsilon_S=2$ (similar to, e.g., polypropylene~\cite{Siag1994}), and a fixed surface charge density $\sigma_S$. The exact charge on a single fibre is difficult to determine: electret filters have an overall low net charge~\cite{Nifuku2001}, meaning that among the individual fibres that compose the filter, there will be similar numbers of positively and negatively charged fibres~\cite{Brown1982}. Consequently, we explore a range of different possibilities, $\sigma_S\in\pm[0.1,1]$~$e/\mathrm{nm}^2$.

\subsection*{Electrostatic interaction in Poisson-Boltzmann approximation}

Even though a virus is likely to reach a respiratory mask while inside of a respiratory droplet, it is less clear what the immediate environment is when it encounters an electret fibre. Salt concentration inside a droplet can vary significantly, and as a droplet shrinks, salt concentration increases---unless efflorescence occurs~\cite{Bozic2020}. Salt concentration can further change once the droplet encounters the electret fibres. We thus consider the cases when the virus and the electret are in an aqueous medium with bulk monovalent 1:1 salt concentrations of $n_0=2$, $10$, and $100$ mM, with the latter value corresponding to physiological salt concentrations such as also encountered in respiratory droplets~\cite{Effros2002,Vejerano2018}.

In the volume of the electret, inside the S proteins, and inside the lipid envelope we have no mobile charges, and the electrostatic potential in these regions satisfies the Poisson equation $\nabla^2\varphi=0$, where $\varphi$ is the electrostatic potential. In all other regions we have an aqueous solution monovalent salt, and the electrostatic potential is governed by the Poisson-Boltzmann (PB) equation~\cite{Safinya}
\begin{equation}
\varepsilon_w \varepsilon_0 \nabla ^2\varphi=2n_0e\sinh{(\beta e\varphi)},
\label{SIPB}
\end{equation}
where $\varepsilon_0$ is the permittivity of vacuum, $\varepsilon_w=78.54$ the dielectric constant of water, and $\beta=1/k_BT$ the inverse thermal energy with $T$ the temperature ($T=298.15$ K in our case) and $k_B$ the Boltzmann constant. The PB equation is accompanied by the condition of continuity of the electrostatic potential everywhere in space, including at charged surfaces and dielectric discontinuities. We also assume that the electrostatic potential is zero infinitely far from the electret and the virus. At all charged surfaces, we employ the standard boundary condition for surface charge density:
\begin{equation}
\sigma(r) =-\varepsilon_w \varepsilon_0\, \frac{\partial \varphi_{out}(r)}{\partial r_\perp}+\varepsilon_w \varepsilon_0 \, \frac{\partial \varphi_{in}(r)}{\partial r_\perp}.
\label{eq:sigma_S_z}
\end{equation}
The surface charge density on the electret is fixed, while the surface charge density on the S proteins is given by Eq.~\eqref{eq:sigma} as a function of the local electrostatic potential.

The PB equation [Eq.~\eqref{SIPB}] with the self-consistent boundary conditions [Eq.~\eqref{eq:sigma_S_z}] are solved using finite-element methods, with the entire system being enclosed  in a cuboid box with bounding surfaces placed by at least a distance of $8/\kappa$ away from the viral surface. The plane $z=0$ is taken as the electret surface with the virus positioned in the top half-space ($z>0$; Fig.~\ref{fig:1}d). We impose periodic boundary conditions in $x$ and $y$ directions. On the top bounding surface of the enclosing box (within the bulk solution), we impose  zero-potential ($\varphi=0$) and, on the bottom surface of the enclosing box (within the electret), we impose zero-charge (zero-field) boundary conditions.

\subsection*{Electrostatic force between virus and electret surface}

On the mean-field level of the PB approximation, the force acting on a body immersed in an electrolyte can be obtained from the integration of the total stress tensor over a closed surface. Written in the Cartesian form, the total stress tensor $T_{ij}=T^E_{ij} + T^O_{ij}$ is the sum of the Maxwell {\em electrostatic} component of the field
\begin{equation}
T^E_{ij} = \varepsilon_0\varepsilon_w\left[\nabla_i\varphi \nabla_j\varphi - {\textstyle\frac{1}{2}}(\bnabla\varphi)^2 \delta_{ij}\right] ,
\end{equation}
and the van't Hoff {\em osmotic} component of the salt ions~\cite{Safinya,Lu_JCP_2005,Gilson_JPC_1993}
\begin{equation}
T^O_{ij} = 2 k_BT n_0 \left[\cosh{(\beta e\varphi)}- 1\right] \delta_{ij}.
\label{eq:osmotic_force} 
\end{equation}
The corresponding {\em interaction force} ${\bf F}$ acting on the virus is obtained by integrating the stress tensor over any surface enclosing the virus:
\begin{equation}
F_i = \oint T_{ij} n_j \mathrm{d}S = F^E_{i} + F^O_{i},
\label{forcetot}
\end{equation}
where $\bf n$ is the outward normal from the surface of the virus and $F^E_{i}$ and $F^O_{i}$ are the electrostatic and osmotic components of the force, respectively, following the decomposition of the stress tensor.

The $z$ coordinate defines the force acting between the virus and the electret: positive or negative $z$ component of the force means repulsion or attraction of the virus to the electret surface, respectively. To calculate this force, we find for each configuration of S proteins three of them nearest to the electret plane and use the plane defined by them as the plane of approach, which also sets the distance $z$ between the virus and the electret (Fig.~\ref{fig:1}d). This choice neglects (numerous) other configurations in which the virus could approach the electret. A full exploration of these configurations is beyond the scope of our work, and the three-protein contact in this sense represents the ``maximum'' contact a virus can have with the electret---both in terms of attraction as well as repulsion.

\section*{Results and Discussion}

\subsection*{Charge regulation modifies charge on spike proteins and their interaction with electret fibre}

To study the electrostatic interaction between a SARS-CoV-2 virus and a charged electret surface, we use a coarse-grained continuum model within the Poisson-Boltzmann (PB) framework conjoined with a charge regulation model describing the protonation-deprotonation reactions of dissociable AAs on the solvent accessible surface (SAS) of the SARS-CoV-2 spike (S) proteins (Models and Methods). We consider both the lipid envelope of the virus as well as the nucleocapsid complex inside it to be uncharged, meaning that only the S proteins contribute to the electrostatic interaction of the virus with the electret fibre. This assumption is supported both by work which shows that electrophoretic mobility~\cite{Johnson1973} and isoelectric points~\cite{Heffron2021} of viruses are not affected by the interior charges as well as by the results presented in this work. The S proteins carry a large number of ionizable AAs (Fig.~\ref{fig:1}a and Table~SI, ESI$\dag$), which impart charge to it through the protonation-deprotonation charging mechanism. In our coarse-grained model we assume that the charge on the S proteins is homogeneously distributed on their SAS in the absence of charge regulation (i.e., without the response of the ionizable AAs to the local electrostatic potential). In this approximation, the S proteins are negatively charged at neutral and basic pH with an isoelectric point of $\mathrm{pI}\approx5.6$ (Fig.~\ref{fig:1}b). When charge regulation is taken into account (Models and Methods), the charge on the S proteins responds to the strength of the local electrostatic potential (shown in an example in Fig.~S1, ESI$\dag$), self-consistently adapting its value and causing in the process the surface charge density on the S proteins to become inhomogeneous (Fig.~\ref{fig:1}c).

Charge regulation has even more significant influence when the virus comes closer to the surface of the electret fibre. As we see in Fig.~\ref{fig:1}d, the proximity of the charged dielectric surface causes the charge on the S proteins closest to it to become extremely inhomogeneous, with charge on the tips of the S proteins even changing sign with respect to the other S proteins further away from the electret. Electrostatic force acting on the virus close to the electret will thus clearly be influenced by charge regulation effects in addition to changes in environmental variables such as pH and salt concentration. After determining the electrostatic potential using a continuum PB model together with charge regulation boundary conditions (Models and Methods), the force between the virus and the electret is obtained from the integral of the stress tensor with additive Maxwell stress and van't Hoff osmotic components, whose contributions to the total force are shown in Fig.~S2 (ESI$\dag$). The net attraction is usually dominated by the Maxwell component, while the Maxwell and the osmotic contributions are more comparable in the case of repulsive interaction. As neither the ionic composition, salt concentration, or the pH in the respiratory droplets are entirely known, especially once they are deposited onto the fibrous electret layer, we need to consistently scan the interaction properties across a range of values that could capture the actual state. We also note that while charge regulation can locally change the sign on the S proteins in the immediate vicinity of the electret fibre (Fig.~\ref{fig:1}d), we are still firmly within the PB theory and interactions between charged molecules with an {\em overall} identical sign of total charge remain always repulsive.

\subsection*{Contact geometry of spike proteins close to electret is as important as their total number}

\begin{figure}[!t]
\centering
\includegraphics[width=0.95\linewidth]{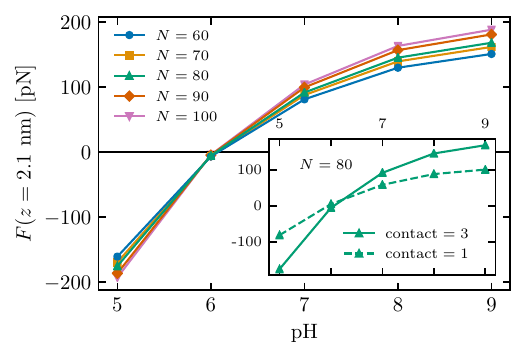}
\caption{Force on the virus at close proximity to the electret surface  ($z = 2.1$ nm) as a function of pH, shown for $n_0=2$ mM, $\sigma_S=-1$ $e_0/\mathrm{nm}^2$, and different numbers of S proteins on the virus $N$. Inset shows the pH dependence of the force for two extreme cases of approach of a virus with $N=80$ S proteins, one where three S proteins are simultaneously in close contact with the electret surface (as in Fig.~\ref{fig:1}d) and one where only one S protein is in close contact with the surface.
}
\label{fig:2}
\end{figure}

To examine first to what extent the total charge on the virus---which in our model corresponds to the number of S proteins it contains---influences the strength of the interaction, we plot in Fig.~\ref{fig:2} the electrostatic force on the virus in close proximity to the electret fibre surface (at a separation comparable to $r_c$, the smallest length scale in our coarse-grained model) as the total number of S proteins on the virus is varied. The virus is oriented in such a way that the plane spanned by three proximal S proteins is parallel to the electret surface (Fig.~\ref{fig:1}d). We see that a larger number of S proteins indeed corresponds to a larger force exerted on the virus---either repulsive or attractive, depending on the pH. The difference between $N=60$ and $N=100$ amounts to between $30$ and $40$ pN difference in force at the extremes of the pH range studied (namely, $\pH=5$ and $\pH=9$).

Interestingly, however, a similar difference in force exerted on the virus is obtained if we simply change the local geometry of the S protein configuration in proximity to the surface of the electret fibre instead of changing their total number. Specifically, as shown in the inset of Fig.~\ref{fig:2}, the difference in force when only a single protein makes close contact with the electret and when three of them make contact simultaneously (for a virus with $N=80$ S proteins) is similar to the difference when $N$ changes from $60$ to $100$. These results indicate that the local geometry of the S proteins close to the electret fibre is at least as important for the electrostatic interaction of the virus with the fibre as their overall number. This is also interesting from the perspective that the S proteins do not necessarily point radially outward from the membrane but can assume also other tilt angles~\cite{Yao2020,Ke2020}, thus influencing the interaction force. Furthermore, if the positions of the S proteins are not fixed in the lipid membrane but can, at least to some extent, move within it, attraction to the fibre could increase the local concentration of the S proteins, increasing the attraction, while repulsion would decrease the local concentration thus also decreasing the repulsion.

These results also further validate our approach, which does not consider the possible charge on the envelope or in the nucleocapsid complex inside it, as this charge is likely even less relevant compared to the contribution of the $N$ S proteins. This conclusion is reinforced by the fact that the surface of the lipid envelope is at a distance of at least two Debye lengths from the electret surface, and even if charged, its contribution to the total interaction would consequently be minimal.

\subsection*{Environmental parameters influence sign and magnitude of interaction force}

\begin{figure}[!t]
\centering
\includegraphics[width=0.95\linewidth]{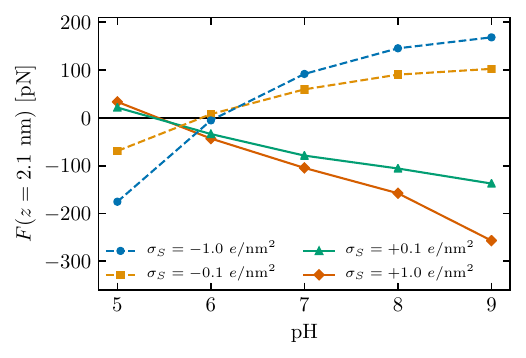}
\caption{Force on the virus at close proximity to the electret surface ($z = 2.1$ nm) as a function of pH, shown for $N=80$, $n_0=2$ mM, and different values of charge density on the electret surface $\sigma_S$.
}
\label{fig:3}
\end{figure}

Figure~\ref{fig:2} also shows that the sign of the force acting on the virus as it approaches the electret surface changes with pH. Whether the force between the virus and the electret fibre is attractive or repulsive thus depends on both the pH---which is one of the determining factors for the overall charge on the virus---as well as the charge on the electret surface. In Fig.~\ref{fig:3}, we see how the sign and the magnitude of the force change with pH as we change the magnitude and sign of the surface charge density on the electret. Interestingly enough, the transition between attraction and repulsion with pH does not happen at the isoelectric point of the S protein as predicted in absence of charge regulation (Fig.~\ref{fig:1}b), but varies depending on the charge on the electret surface. This indicates that the isoelectric point of the protein is not a fixed value but can change as a consequence of charge regulation, which can also be due to the presence of other charged objects in its vicinity. In addition, we can observe a notable asymmetry in the interaction strength between positively and negatively charged electret surfaces at identical absolute value of the surface charge. At the same solution conditions, attractive force tends to be larger than the corresponding repulsive force, with the difference increasing towards more extreme values of pH. What is more, when the repulsive force vanishes around $pH\approx6$, the attractive force remains non-zero, and this environment could thus provide better conditions for SARS-CoV-2 virus adsorption.

\begin{figure}[!htp]
\centering
\includegraphics[width=0.95\linewidth]{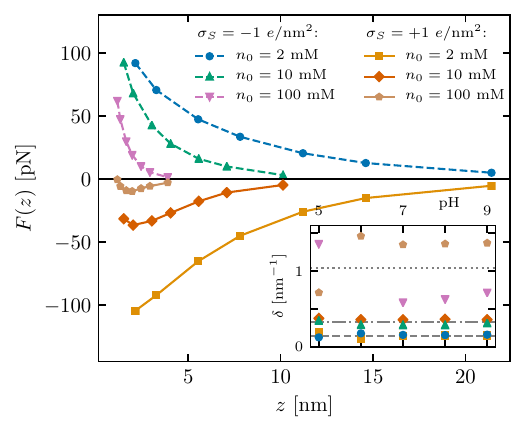}
\caption{Force on the virus as a function of distance from the electret surface, shown for $N=80$, $\pH=7$, and different values of salt concentration $n_0$ and different signs of charge density on the electret surface $\sigma_S$. Inset shows the inverse decay length of the force at large separations at different pH, obtained from a fit of the form $F(z)=a\exp(-\delta z)$, and the horizontal lines show the inverse Debye lengths $\kappa$ at $n_0=2$ mM (dashed), $10$ mM (dot-dashed), and $100$ mM (dotted).
}
\label{fig:4}
\end{figure}

While pH and the surface charge density on the electret fibre together determine whether the force on the virus will be attractive or repulsive, the range of the interaction is mainly determined by the screening due to the salt ions in the surrounding medium. Figure~\ref{fig:4} shows how the electrostatic force decays with distance from the electret for different values of salt concentration $n_0$. As one would expect, the interaction is stronger in the case of smaller $n_0$ because of weaker screening. At larger distances, the force decays with the Debye screening length $\kappa^{-1}=(8 \pi \ell_B n_0)^{-1/2}$ (with $\ell_B=e^2/4\pi\epsilon\epsilon_0 k_B T$), as shown in the inset of Fig.~\ref{fig:4}. Fitting the long-range decay of the force with $F(z)=a\exp(-\delta z)$ and comparing the exponent with the Debye screening length, it is interesting to observe that the decay length of the force deviates from the Debye length the most when the salt concentration is high. The reason for this could be ascribed to geometric details of the corrugated virion geometry that are not smoothed out by the electric field with a small Debye screening length, and their effect persists to separations on the order of the corrugation scale. Asymptotically, of course, the screening length will reverse to the Debye length~\cite{cats2021}.

\begin{figure}[!htp]
\centering
\includegraphics[width=0.95\linewidth]{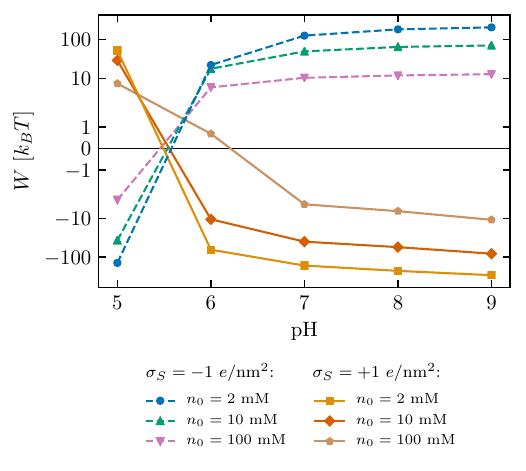}
\caption{
Interaction energy between the virus and the electret surface, obtained from numerical integration of the interaction force [Eq.~\eqref{forcetot}] between $z = 3 \kappa^{-1}$ and $z = 1 \textrm{ nm} + \kappa^{-1}/6$. The largest absolute values of this interaction energy, on the order of $\simeq 100 {\rm k_B T}$, are obtained for extreme acidic or basic values of the pH when the salt concentration is low, $n_0=2$ mM.
}
\label{fig:5}
\end{figure}

The interaction energy between the virus and the electret is obtained from the work needed to bring  the virus to the electret surface, i.e., from the numerical integration of the interaction force [Eq.~\eqref{forcetot}] with respect to the distance between the two. The distance of closest approach is taken as $z = 1 \textrm{ nm} + \kappa^{-1}/6$.  The interaction energy is sensitive to both pH and salt concentration, varying from $\simeq10$ $k_BT$ in the case of $n_0=100$ mM to $\simeq100$ $k_BT$ when $n_0=2$ mM (Fig.~\ref{fig:5}).

\subsection*{Contribution of other long-range interactions}

Apart from the mean-field PB electrostatics, the classical Deryaguin-Landau-Verwey-Overbeek (DLVO) theory of colloidal stability also implies the electrodynamic van der Waals (vdW) interactions that could make a contribution to the virion--electret interactions at the smallest values of their separation. Just as in the electrostatic case, the largest contribution to the vdW interactions would stem from the S protein--electret interactions. At the separation of a few nanometres and assuming a reasonable value for the Hamaker coefficient~\cite{Lenhoff1996}, one obtains an upper bound for the vdW force $\sim50$ pN at $1$ nm separation of the S protein from the electret surface, which is too small to qualitatively change the interaction picture that we drew here. In addition, at small separations there would be interaction contributions from a host of other, non-DLVO mechanisms, which are presently difficult to quantify~\cite{Embarras2020}.

\section*{Conclusions}

As a first insight into the details of the interaction between a single charged virus particle and a single charged electret fibre, we have analyzed the details of the electrostatic interaction between a SARS-CoV-2 continuum charge model with explicit S protein geometry and a surface of a single charged polypropylene electret fibre. Given the many uncertainties in the details of the actual bathing solution environment, we scanned the parameter space of the bathing solution within a range of plausible values. We used the same general approach for the electret fibre surface, as fibres with both signs of charge in the macroscopic electret filters sample are possible due to the charging process used to generate them. As we clearly demonstrate, these indeterminacies should not be seen as impediments as they actually allow for the electret filters to successfully capture different viruses across a broad range of species and environments.

Our results show that a model of the S proteins that explicitly takes into account the AA dissociation equilibrium on the solvent accessible surface enables the description of changes in the protein charge state on approach to the electret surface and is crucial for the quantification of the electrostatic interaction. Namely, we observe that the charge on the S proteins closest to the electret fibre depends on the separation from the electret surface and can even reverse sign. The calculated magnitude of the electrostatic force also allows us to pinpoint the importance of the contact geometry of the S proteins, which to a large extent sets the scale of the interaction and is as important as the total charge of the S proteins contained on the virus. Furthermore, the environmental pH, solution conditions, and the charge of the electret fibre surface together set the sign and magnitude of the interaction, delineating environmental conditions which are potentially better suited for SARS-CoV-2 adsorption.

In our work, we have focused solely on the contribution of the S proteins to the charge on the virus and thus to its electrostatic interaction with the electret. While AFM experiments have measured a difference in the electrostatic characteristics of empty and genome-filled capsids of some viruses~\cite{Hernando2015}, other work on electrophoretic mobility~\cite{Johnson1973} and isoelectric points of different viruses~\cite{Heffron2021} indicates that this might not be the case, and that the outermost charges dominate the electrostatic characteristics of a virus. That the latter might be the case is indicated also by the results of our work; however, a dedicated theoretical study would be needed to address this issue. Things are further complicated by the complex structure of the lipid envelope of SARS-CoV-2, which includes several membrane proteins and potentially charged lipids, and the essentially unknown structure of the nucleocapsid complex inside it. The model we use is flexible and can incorporate additional structural details of SARS-CoV-2 as they become available, and can be easily extended to other types of viruses as well. Nevertheless, because of the observed overwhelming importance of the charge state and interaction geometry of S proteins proximal to the electret surface, even the present level of structural detail that does not explicitly incorporate the membrane (protein) charge or the charge of the nucleocapsid complex seems to be able to describe the electrostatics of SARS-CoV-2 and electret fibre surface interaction to quantitative detail.

\section*{Author contributions}
All authors developed the model, analyzed the data and wrote the manuscript. LJ performed the numerical computations. RP and AN acted as principal investigators. 

\section*{Conflicts of interest}
There are no conflicts to declare.

\section*{Acknowledgements}
AB acknowledges funding from the Slovenian Research Agency ARRS (Research Core Funding No.\ P1-0055 and research Grant No.\ J1-9149). RP acknowledges funding from the Key project \#12034019 of the National Natural Science Foundation of China. AN and LJ acknowledge partial support from the Iran National Science Foundation (grant No.\ 98022853). AN acknowledges partial support from the Associateship Scheme of The Abdus Salam International Centre for Theoretical Physics (Trieste, Italy). AN thanks the School of Physics, University of Chinese Academy of Sciences, Beijing, for hospitality and travel support.

\balance

\bibliography{rsc} 

\providecommand*{\mcitethebibliography}{\thebibliography}
\csname @ifundefined\endcsname{endmcitethebibliography}
{\let\endmcitethebibliography\endthebibliography}{}
\begin{mcitethebibliography}{64}
\providecommand*{\natexlab}[1]{#1}
\providecommand*{\mciteSetBstSublistMode}[1]{}
\providecommand*{\mciteSetBstMaxWidthForm}[2]{}
\providecommand*{\mciteBstWouldAddEndPuncttrue}
  {\def\EndOfBibitem{\unskip.}}
\providecommand*{\mciteBstWouldAddEndPunctfalse}
  {\let\EndOfBibitem\relax}
\providecommand*{\mciteSetBstMidEndSepPunct}[3]{}
\providecommand*{\mciteSetBstSublistLabelBeginEnd}[3]{}
\providecommand*{\EndOfBibitem}{}
\mciteSetBstSublistMode{f}
\mciteSetBstMaxWidthForm{subitem}
{(\emph{\alph{mcitesubitemcount}})}
\mciteSetBstSublistLabelBeginEnd{\mcitemaxwidthsubitemform\space}
{\relax}{\relax}

\bibitem[Kutter \emph{et~al.}(2018)Kutter, Spronken, Fraaij, Fouchier, and
  Herfst]{Kutter2018}
J.~S. Kutter, M.~I. Spronken, P.~L. Fraaij, R.~A. Fouchier and S.~Herfst,
  \emph{Curr. Op. Virol.}, 2018, \textbf{28}, 142--151\relax
\mciteBstWouldAddEndPuncttrue
\mciteSetBstMidEndSepPunct{\mcitedefaultmidpunct}
{\mcitedefaultendpunct}{\mcitedefaultseppunct}\relax
\EndOfBibitem
\bibitem[Fernstrom and Goldblatt(2013)]{Fernstrom2013}
A.~Fernstrom and M.~Goldblatt, \emph{J. Pathog.}, 2013, \textbf{2013},
  493960\relax
\mciteBstWouldAddEndPuncttrue
\mciteSetBstMidEndSepPunct{\mcitedefaultmidpunct}
{\mcitedefaultendpunct}{\mcitedefaultseppunct}\relax
\EndOfBibitem
\bibitem[La~Rosa \emph{et~al.}(2013)La~Rosa, Fratini, Libera, Iaconelli, and
  Muscillo]{laRosa2013}
G.~La~Rosa, M.~Fratini, S.~D. Libera, M.~Iaconelli and M.~Muscillo, \emph{Ann.
  Ist. Super. Sanita}, 2013, \textbf{49}, 124--132\relax
\mciteBstWouldAddEndPuncttrue
\mciteSetBstMidEndSepPunct{\mcitedefaultmidpunct}
{\mcitedefaultendpunct}{\mcitedefaultseppunct}\relax
\EndOfBibitem
\bibitem[{Bo\v{z}i\v{c}} and Kandu\v{c}(2021)]{Bozic2020}
A.~{Bo\v{z}i\v{c}} and M.~Kandu\v{c}, \emph{J. Biol. Phys.}, 2021,  1--29\relax
\mciteBstWouldAddEndPuncttrue
\mciteSetBstMidEndSepPunct{\mcitedefaultmidpunct}
{\mcitedefaultendpunct}{\mcitedefaultseppunct}\relax
\EndOfBibitem
\bibitem[Dancer \emph{et~al.}(2020)Dancer, Tang, Marr, Miller, Morawska, and
  Jimenez]{Dancer2020}
S.~J. Dancer, J.~W. Tang, L.~C. Marr, S.~Miller, L.~Morawska and J.~L. Jimenez,
  \emph{J. Hosp. Infect.}, 2020, \textbf{105}, 569--570\relax
\mciteBstWouldAddEndPuncttrue
\mciteSetBstMidEndSepPunct{\mcitedefaultmidpunct}
{\mcitedefaultendpunct}{\mcitedefaultseppunct}\relax
\EndOfBibitem
\bibitem[Prather \emph{et~al.}(2020)Prather, Wang, and Schooley]{Prather2020}
K.~A. Prather, C.~C. Wang and R.~T. Schooley, \emph{Science}, 2020,
  \textbf{368}, 1422--1424\relax
\mciteBstWouldAddEndPuncttrue
\mciteSetBstMidEndSepPunct{\mcitedefaultmidpunct}
{\mcitedefaultendpunct}{\mcitedefaultseppunct}\relax
\EndOfBibitem
\bibitem[Morawska \emph{et~al.}(2020)Morawska, Tang, Bahnfleth, Bluyssen,
  Boerstra, Buonanno, Cao, Dancer, Floto,
  Franchimon,\emph{et~al.}]{Morawska2020a}
L.~Morawska, J.~W. Tang, W.~Bahnfleth, P.~M. Bluyssen, A.~Boerstra,
  G.~Buonanno, J.~Cao, S.~Dancer, A.~Floto, F.~Franchimon \emph{et~al.},
  \emph{Environ. Int.}, 2020, \textbf{142}, 105832\relax
\mciteBstWouldAddEndPuncttrue
\mciteSetBstMidEndSepPunct{\mcitedefaultmidpunct}
{\mcitedefaultendpunct}{\mcitedefaultseppunct}\relax
\EndOfBibitem
\bibitem[Chu \emph{et~al.}(2020)Chu, Akl, Duda, Solo, Yaacoub, and
  Sch{\"u}nemann]{Chu2020}
D.~K. Chu, E.~A. Akl, S.~Duda, K.~Solo, S.~Yaacoub and H.~J. Sch{\"u}nemann,
  \emph{Lancet}, 2020, \textbf{395}, 1973--1987\relax
\mciteBstWouldAddEndPuncttrue
\mciteSetBstMidEndSepPunct{\mcitedefaultmidpunct}
{\mcitedefaultendpunct}{\mcitedefaultseppunct}\relax
\EndOfBibitem
\bibitem[Gandhi \emph{et~al.}(2020)Gandhi, Beyrer, and Goosby]{Gandhi2020}
M.~Gandhi, C.~Beyrer and E.~Goosby, \emph{J. Gen. Intern. Med.}, 2020,
  \textbf{35}, 3063--3066\relax
\mciteBstWouldAddEndPuncttrue
\mciteSetBstMidEndSepPunct{\mcitedefaultmidpunct}
{\mcitedefaultendpunct}{\mcitedefaultseppunct}\relax
\EndOfBibitem
\bibitem[Davies \emph{et~al.}(2013)Davies, Thompson, Giri, Kafatos, Walker, and
  Bennett]{Davies2013}
A.~Davies, K.-A. Thompson, K.~Giri, G.~Kafatos, J.~Walker and A.~Bennett,
  \emph{Disaster Med Public Health Prep.}, 2013, \textbf{7}, 413--418\relax
\mciteBstWouldAddEndPuncttrue
\mciteSetBstMidEndSepPunct{\mcitedefaultmidpunct}
{\mcitedefaultendpunct}{\mcitedefaultseppunct}\relax
\EndOfBibitem
\bibitem[Fischer \emph{et~al.}(2020)Fischer, Fischer, Grass, Henrion, Warren,
  and Westman]{Fischer2020}
E.~P. Fischer, M.~C. Fischer, D.~Grass, I.~Henrion, W.~S. Warren and
  E.~Westman, \emph{Sci. Adv.}, 2020, \textbf{6}, eabd3083\relax
\mciteBstWouldAddEndPuncttrue
\mciteSetBstMidEndSepPunct{\mcitedefaultmidpunct}
{\mcitedefaultendpunct}{\mcitedefaultseppunct}\relax
\EndOfBibitem
\bibitem[Long \emph{et~al.}(2020)Long, Hu, Liu, Chen, Guo, Yang, Cheng, Huang,
  and Du]{Long2020}
Y.~Long, T.~Hu, L.~Liu, R.~Chen, Q.~Guo, L.~Yang, Y.~Cheng, J.~Huang and L.~Du,
  \emph{J. Evid. Based Med.}, 2020, \textbf{13}, 93--101\relax
\mciteBstWouldAddEndPuncttrue
\mciteSetBstMidEndSepPunct{\mcitedefaultmidpunct}
{\mcitedefaultendpunct}{\mcitedefaultseppunct}\relax
\EndOfBibitem
\bibitem[Liao \emph{et~al.}(2020)Liao, Xiao, Zhao, Yu, Wang, Wang, Chu, and
  Cui]{Liao2020}
L.~Liao, W.~Xiao, M.~Zhao, X.~Yu, H.~Wang, Q.~Wang, S.~Chu and Y.~Cui,
  \emph{ACS Nano}, 2020, \textbf{14}, 6348--6356\relax
\mciteBstWouldAddEndPuncttrue
\mciteSetBstMidEndSepPunct{\mcitedefaultmidpunct}
{\mcitedefaultendpunct}{\mcitedefaultseppunct}\relax
\EndOfBibitem
\bibitem[Dyrud \emph{et~al.}(1989)Dyrud, Berg, and Murray]{Dyrud1989}
J.~F. Dyrud, H.~J. Berg and A.~C. Murray, \emph{Resilient shape-retaining
  fibrous filtration face mask}, 1989, {US} {Patent} 4,807,619\relax
\mciteBstWouldAddEndPuncttrue
\mciteSetBstMidEndSepPunct{\mcitedefaultmidpunct}
{\mcitedefaultendpunct}{\mcitedefaultseppunct}\relax
\EndOfBibitem
\bibitem[Hossain \emph{et~al.}(2020)Hossain, Bhadra, Jain, Das, Bhattacharya,
  Ghosh, and Levine]{Hossain2020}
E.~Hossain, S.~Bhadra, H.~Jain, S.~Das, A.~Bhattacharya, S.~Ghosh and
  D.~Levine, \emph{Phys. Fluids}, 2020, \textbf{32}, 093304\relax
\mciteBstWouldAddEndPuncttrue
\mciteSetBstMidEndSepPunct{\mcitedefaultmidpunct}
{\mcitedefaultendpunct}{\mcitedefaultseppunct}\relax
\EndOfBibitem
\bibitem[Wang(2001)]{Wang2001}
C.-S. Wang, \emph{Powder Technol.}, 2001, \textbf{118}, 166--170\relax
\mciteBstWouldAddEndPuncttrue
\mciteSetBstMidEndSepPunct{\mcitedefaultmidpunct}
{\mcitedefaultendpunct}{\mcitedefaultseppunct}\relax
\EndOfBibitem
\bibitem[Lundgren and Whitby(1965)]{Lundgren1965}
D.~A. Lundgren and K.~Whitby, \emph{Ind. Eng. Chem.}, 1965, \textbf{4},
  345--349\relax
\mciteBstWouldAddEndPuncttrue
\mciteSetBstMidEndSepPunct{\mcitedefaultmidpunct}
{\mcitedefaultendpunct}{\mcitedefaultseppunct}\relax
\EndOfBibitem
\bibitem[Fjeld and Owens(1988)]{Fjeld1988}
R.~A. Fjeld and T.~M. Owens, \emph{IEEE Trans. Ind. Appl.}, 1988, \textbf{24},
  725--731\relax
\mciteBstWouldAddEndPuncttrue
\mciteSetBstMidEndSepPunct{\mcitedefaultmidpunct}
{\mcitedefaultendpunct}{\mcitedefaultseppunct}\relax
\EndOfBibitem
\bibitem[Romay \emph{et~al.}(1998)Romay, Liu, and Chae]{Romay1998}
F.~J. Romay, B.~Y. Liu and S.-J. Chae, \emph{Aerosol Sci. Technol.}, 1998,
  \textbf{28}, 224--234\relax
\mciteBstWouldAddEndPuncttrue
\mciteSetBstMidEndSepPunct{\mcitedefaultmidpunct}
{\mcitedefaultendpunct}{\mcitedefaultseppunct}\relax
\EndOfBibitem
\bibitem[Ardkapan \emph{et~al.}(2014)Ardkapan, Johnson, Yazdi, Afshari, and
  Bergs{\o}e]{Ardkapan2014}
S.~R. Ardkapan, M.~S. Johnson, S.~Yazdi, A.~Afshari and N.~C. Bergs{\o}e,
  \emph{J. Aerosol Sci.}, 2014, \textbf{72}, 14--20\relax
\mciteBstWouldAddEndPuncttrue
\mciteSetBstMidEndSepPunct{\mcitedefaultmidpunct}
{\mcitedefaultendpunct}{\mcitedefaultseppunct}\relax
\EndOfBibitem
\bibitem[Chen and Huang(1998)]{Chen1998}
C.-C. Chen and S.-H. Huang, \emph{Am. Ind. Hyg. Assoc. J.}, 1998, \textbf{59},
  227--233\relax
\mciteBstWouldAddEndPuncttrue
\mciteSetBstMidEndSepPunct{\mcitedefaultmidpunct}
{\mcitedefaultendpunct}{\mcitedefaultseppunct}\relax
\EndOfBibitem
\bibitem[Tsai \emph{et~al.}(2002)Tsai, Schreuder-Gibson, and Gibson]{Tsai2002}
P.~P. Tsai, H.~Schreuder-Gibson and P.~Gibson, \emph{J. Electrostat.}, 2002,
  \textbf{54}, 333--341\relax
\mciteBstWouldAddEndPuncttrue
\mciteSetBstMidEndSepPunct{\mcitedefaultmidpunct}
{\mcitedefaultendpunct}{\mcitedefaultseppunct}\relax
\EndOfBibitem
\bibitem[Walsh and Stenhouse(1997)]{Walsh1997}
D.~Walsh and J.~Stenhouse, \emph{Powder Technol.}, 1997, \textbf{93},
  63--75\relax
\mciteBstWouldAddEndPuncttrue
\mciteSetBstMidEndSepPunct{\mcitedefaultmidpunct}
{\mcitedefaultendpunct}{\mcitedefaultseppunct}\relax
\EndOfBibitem
\bibitem[Kraemer and Johnstone(1955)]{Kraemer1955}
H.~F. Kraemer and H.~Johnstone, \emph{Ind. Eng. Chem.}, 1955, \textbf{47},
  2426--2434\relax
\mciteBstWouldAddEndPuncttrue
\mciteSetBstMidEndSepPunct{\mcitedefaultmidpunct}
{\mcitedefaultendpunct}{\mcitedefaultseppunct}\relax
\EndOfBibitem
\bibitem[Lathrache and Fissan(1987)]{Lathrache1987}
R.~Lathrache and H.~Fissan, \emph{Filtr. Sep.}, 1987, \textbf{24},
  418--422\relax
\mciteBstWouldAddEndPuncttrue
\mciteSetBstMidEndSepPunct{\mcitedefaultmidpunct}
{\mcitedefaultendpunct}{\mcitedefaultseppunct}\relax
\EndOfBibitem
\bibitem[Thakur \emph{et~al.}(2013)Thakur, Das, and Das]{Thakur2013}
R.~Thakur, D.~Das and A.~Das, \emph{Sep. Purif. Rev.}, 2013, \textbf{42},
  87--129\relax
\mciteBstWouldAddEndPuncttrue
\mciteSetBstMidEndSepPunct{\mcitedefaultmidpunct}
{\mcitedefaultendpunct}{\mcitedefaultseppunct}\relax
\EndOfBibitem
\bibitem[{Bo\v{z}i\v{c}} \emph{et~al.}(2012){Bo\v{z}i\v{c}}, \v{S}iber, and
  Podgornik]{Bozic2012}
A.~{Bo\v{z}i\v{c}}, A.~\v{S}iber and R.~Podgornik, \emph{J. Biol. Phys.}, 2012,
  \textbf{38}, 657--671\relax
\mciteBstWouldAddEndPuncttrue
\mciteSetBstMidEndSepPunct{\mcitedefaultmidpunct}
{\mcitedefaultendpunct}{\mcitedefaultseppunct}\relax
\EndOfBibitem
\bibitem[{Bo\v{z}i\v{c}} and Podgornik(2017)]{Bozic2017b}
A.~{Bo\v{z}i\v{c}} and R.~Podgornik, \emph{J. Phys. Condens. Matter}, 2017,
  \textbf{30}, 024001\relax
\mciteBstWouldAddEndPuncttrue
\mciteSetBstMidEndSepPunct{\mcitedefaultmidpunct}
{\mcitedefaultendpunct}{\mcitedefaultseppunct}\relax
\EndOfBibitem
\bibitem[Bo\v{z}i\v{c} and Podgornik(2017)]{Bozic2017}
A.~Bo\v{z}i\v{c} and R.~Podgornik, \emph{Biophys. J.}, 2017, \textbf{113},
  1454--1465\relax
\mciteBstWouldAddEndPuncttrue
\mciteSetBstMidEndSepPunct{\mcitedefaultmidpunct}
{\mcitedefaultendpunct}{\mcitedefaultseppunct}\relax
\EndOfBibitem
\bibitem[Bo\v{z}i\v{c} and Podgornik(2018)]{Bozic2018}
A.~Bo\v{z}i\v{c} and R.~Podgornik, \emph{J. Chem. Phys.}, 2018, \textbf{149},
  1163307\relax
\mciteBstWouldAddEndPuncttrue
\mciteSetBstMidEndSepPunct{\mcitedefaultmidpunct}
{\mcitedefaultendpunct}{\mcitedefaultseppunct}\relax
\EndOfBibitem
\bibitem[Nap \emph{et~al.}(2014)Nap, Bo\v{z}i\v{c}, Szleifer, and
  Podgornik]{Nap_Anze_2014}
R.~J. Nap, A.~Bo\v{z}i\v{c}, I.~Szleifer and R.~Podgornik, \emph{Biophys. J.},
  2014, \textbf{107}, 1970--1979\relax
\mciteBstWouldAddEndPuncttrue
\mciteSetBstMidEndSepPunct{\mcitedefaultmidpunct}
{\mcitedefaultendpunct}{\mcitedefaultseppunct}\relax
\EndOfBibitem
\bibitem[Markovich \emph{et~al.}(2015)Markovich, Andelman, and
  Podgornik]{Markovich2015}
T.~Markovich, D.~Andelman and R.~Podgornik, \emph{J. Chem. Phys.}, 2015,
  \textbf{142}, 044702\relax
\mciteBstWouldAddEndPuncttrue
\mciteSetBstMidEndSepPunct{\mcitedefaultmidpunct}
{\mcitedefaultendpunct}{\mcitedefaultseppunct}\relax
\EndOfBibitem
\bibitem[Podgornik(2018)]{GeneralCR2019}
R.~Podgornik, \emph{J. Chem. Phys.}, 2018, \textbf{149}, 104701\relax
\mciteBstWouldAddEndPuncttrue
\mciteSetBstMidEndSepPunct{\mcitedefaultmidpunct}
{\mcitedefaultendpunct}{\mcitedefaultseppunct}\relax
\EndOfBibitem
\bibitem[Krishnan(2017)]{Krishnan2017}
M.~Krishnan, \emph{J. Chem. Phys.}, 2017, \textbf{146}, 205101\relax
\mciteBstWouldAddEndPuncttrue
\mciteSetBstMidEndSepPunct{\mcitedefaultmidpunct}
{\mcitedefaultendpunct}{\mcitedefaultseppunct}\relax
\EndOfBibitem
\bibitem[Atalay \emph{et~al.}(2014)Atalay, Barisik, Beskok, and
  Qian]{Atalay2014}
S.~Atalay, M.~Barisik, A.~Beskok and S.~Qian, \emph{J. Phys. Chem. C}, 2014,
  \textbf{118}, 10927--10935\relax
\mciteBstWouldAddEndPuncttrue
\mciteSetBstMidEndSepPunct{\mcitedefaultmidpunct}
{\mcitedefaultendpunct}{\mcitedefaultseppunct}\relax
\EndOfBibitem
\bibitem[Brown(1982)]{Brown1982}
R.~Brown, \emph{J. Aerosol Sci.}, 1982, \textbf{13}, 249--257\relax
\mciteBstWouldAddEndPuncttrue
\mciteSetBstMidEndSepPunct{\mcitedefaultmidpunct}
{\mcitedefaultendpunct}{\mcitedefaultseppunct}\relax
\EndOfBibitem
\bibitem[Nifuku \emph{et~al.}(2001)Nifuku, Zhou, Kisiel, Kobayashi, and
  Katoh]{Nifuku2001}
M.~Nifuku, Y.~Zhou, A.~Kisiel, T.~Kobayashi and H.~Katoh, \emph{J.
  Electrostat.}, 2001, \textbf{51}, 200--205\relax
\mciteBstWouldAddEndPuncttrue
\mciteSetBstMidEndSepPunct{\mcitedefaultmidpunct}
{\mcitedefaultendpunct}{\mcitedefaultseppunct}\relax
\EndOfBibitem
\bibitem[Ke \emph{et~al.}(2020)Ke, Oton, Qu, Cortese, Zila, McKeane, Nakane,
  Zivanov, Neufeldt, Cerikan,\emph{et~al.}]{Ke2020}
Z.~Ke, J.~Oton, K.~Qu, M.~Cortese, V.~Zila, L.~McKeane, T.~Nakane, J.~Zivanov,
  C.~J. Neufeldt, B.~Cerikan \emph{et~al.}, \emph{Nature}, 2020, \textbf{588},
  498--502\relax
\mciteBstWouldAddEndPuncttrue
\mciteSetBstMidEndSepPunct{\mcitedefaultmidpunct}
{\mcitedefaultendpunct}{\mcitedefaultseppunct}\relax
\EndOfBibitem
\bibitem[Yao \emph{et~al.}(2020)Yao, Song, Chen, Wu, Xu, Sun, Zhang, Weng,
  Zhang, Wu,\emph{et~al.}]{Yao2020}
H.~Yao, Y.~Song, Y.~Chen, N.~Wu, J.~Xu, C.~Sun, J.~Zhang, T.~Weng, Z.~Zhang,
  Z.~Wu \emph{et~al.}, \emph{Cell}, 2020, \textbf{183}, 730--738\relax
\mciteBstWouldAddEndPuncttrue
\mciteSetBstMidEndSepPunct{\mcitedefaultmidpunct}
{\mcitedefaultendpunct}{\mcitedefaultseppunct}\relax
\EndOfBibitem
\bibitem[Bar-On \emph{et~al.}(2020)Bar-On, Flamholz, Phillips, and
  Milo]{Bar2020}
Y.~M. Bar-On, A.~Flamholz, R.~Phillips and R.~Milo, \emph{eLife}, 2020,
  \textbf{9}, e57309\relax
\mciteBstWouldAddEndPuncttrue
\mciteSetBstMidEndSepPunct{\mcitedefaultmidpunct}
{\mcitedefaultendpunct}{\mcitedefaultseppunct}\relax
\EndOfBibitem
\bibitem[Beniac \emph{et~al.}(2006)Beniac, Andonov, Grudeski, and
  Booth]{Beniac2006}
D.~R. Beniac, A.~Andonov, E.~Grudeski and T.~F. Booth, \emph{Nature Struct.
  Mol. Biol.}, 2006, \textbf{13}, 751--752\relax
\mciteBstWouldAddEndPuncttrue
\mciteSetBstMidEndSepPunct{\mcitedefaultmidpunct}
{\mcitedefaultendpunct}{\mcitedefaultseppunct}\relax
\EndOfBibitem
\bibitem[Gramse \emph{et~al.}(2013)Gramse, Dols-Perez, Edwards, Fumagalli, and
  Gomila]{Gramse2013}
G.~Gramse, A.~Dols-Perez, M.~Edwards, L.~Fumagalli and G.~Gomila, \emph{Biophys
  J.}, 2013, \textbf{104}, 1257--1262\relax
\mciteBstWouldAddEndPuncttrue
\mciteSetBstMidEndSepPunct{\mcitedefaultmidpunct}
{\mcitedefaultendpunct}{\mcitedefaultseppunct}\relax
\EndOfBibitem
\bibitem[Walls \emph{et~al.}(2020)Walls, Park, Tortorici, Wall, Mcguire, and
  Veesler]{Walls2020}
A.~Walls, Y.-J. Park, M.~A. Tortorici, A.~Wall, A.~Mcguire and D.~Veesler,
  \emph{Cell}, 2020, \textbf{181}, 281--292\relax
\mciteBstWouldAddEndPuncttrue
\mciteSetBstMidEndSepPunct{\mcitedefaultmidpunct}
{\mcitedefaultendpunct}{\mcitedefaultseppunct}\relax
\EndOfBibitem
\bibitem[{\v{S}}iber(2020)]{Siber2020}
A.~{\v{S}}iber, \emph{Symmetry}, 2020, \textbf{12}, 556\relax
\mciteBstWouldAddEndPuncttrue
\mciteSetBstMidEndSepPunct{\mcitedefaultmidpunct}
{\mcitedefaultendpunct}{\mcitedefaultseppunct}\relax
\EndOfBibitem
\bibitem[Zandi \emph{et~al.}(2020)Zandi, Dragnea, Travesset, and
  Podgornik]{Zandi2020}
R.~Zandi, B.~Dragnea, A.~Travesset and R.~Podgornik, \emph{Phys. Rep.}, 2020,
  \textbf{847}, 1 -- 102\relax
\mciteBstWouldAddEndPuncttrue
\mciteSetBstMidEndSepPunct{\mcitedefaultmidpunct}
{\mcitedefaultendpunct}{\mcitedefaultseppunct}\relax
\EndOfBibitem
\bibitem[Neuman \emph{et~al.}(2006)Neuman, Adair, Yoshioka, Quispe, Orca, Kuhn,
  Milligan, Yeager, and Buchmeier]{Neuman2006}
B.~W. Neuman, B.~D. Adair, C.~Yoshioka, J.~D. Quispe, G.~Orca, P.~Kuhn, R.~A.
  Milligan, M.~Yeager and M.~J. Buchmeier, \emph{J. Virol.}, 2006, \textbf{80},
  7918--7928\relax
\mciteBstWouldAddEndPuncttrue
\mciteSetBstMidEndSepPunct{\mcitedefaultmidpunct}
{\mcitedefaultendpunct}{\mcitedefaultseppunct}\relax
\EndOfBibitem
\bibitem[Neuman \emph{et~al.}(2011)Neuman, Kiss, Kunding, Bhella, Baksh,
  Connelly, Droese, Klaus, Makino, Sawicki,\emph{et~al.}]{Neuman2011}
B.~W. Neuman, G.~Kiss, A.~H. Kunding, D.~Bhella, M.~F. Baksh, S.~Connelly,
  B.~Droese, J.~P. Klaus, S.~Makino, S.~G. Sawicki \emph{et~al.}, \emph{J.
  Struct. Biol.}, 2011, \textbf{174}, 11--22\relax
\mciteBstWouldAddEndPuncttrue
\mciteSetBstMidEndSepPunct{\mcitedefaultmidpunct}
{\mcitedefaultendpunct}{\mcitedefaultseppunct}\relax
\EndOfBibitem
\bibitem[Mitchell(1991)]{Mitchell1991}
D.~P. Mitchell, SIGGRAPH Comput. Graph., 1991, pp. 157--164\relax
\mciteBstWouldAddEndPuncttrue
\mciteSetBstMidEndSepPunct{\mcitedefaultmidpunct}
{\mcitedefaultendpunct}{\mcitedefaultseppunct}\relax
\EndOfBibitem
\bibitem[Bo{\v{z}}i{\v{c}} and {\v{C}}opar(2019)]{Bozic2019}
A.~Bo{\v{z}}i{\v{c}} and S.~{\v{C}}opar, \emph{Phys. Rev. E}, 2019,
  \textbf{99}, 032601\relax
\mciteBstWouldAddEndPuncttrue
\mciteSetBstMidEndSepPunct{\mcitedefaultmidpunct}
{\mcitedefaultendpunct}{\mcitedefaultseppunct}\relax
\EndOfBibitem
\bibitem[Jim{\'e}nez-Zaragoza \emph{et~al.}(2018)Jim{\'e}nez-Zaragoza, Yubero,
  Mart{\'\i}n-Forero, Cast{\'o}n, Reguera, Luque, de~Pablo, and
  Rodr{\'\i}guez]{Jimenez2018}
M.~Jim{\'e}nez-Zaragoza, M.~P. Yubero, E.~Mart{\'\i}n-Forero, J.~R. Cast{\'o}n,
  D.~Reguera, D.~Luque, P.~J. de~Pablo and J.~M. Rodr{\'\i}guez, \emph{eLife},
  2018, \textbf{7}, e37295\relax
\mciteBstWouldAddEndPuncttrue
\mciteSetBstMidEndSepPunct{\mcitedefaultmidpunct}
{\mcitedefaultendpunct}{\mcitedefaultseppunct}\relax
\EndOfBibitem
\bibitem[Adhikari \emph{et~al.}(2020)Adhikari, Li, Shin, Steinmetz, Twarock,
  Podgornik, and Ching]{Adhikari2020}
P.~Adhikari, N.~Li, M.~Shin, N.~F. Steinmetz, R.~Twarock, R.~Podgornik and
  W.-Y. Ching, \emph{Phys. Chem. Chem. Phys.}, 2020, \textbf{22},
  18272--18283\relax
\mciteBstWouldAddEndPuncttrue
\mciteSetBstMidEndSepPunct{\mcitedefaultmidpunct}
{\mcitedefaultendpunct}{\mcitedefaultseppunct}\relax
\EndOfBibitem
\bibitem[Siag \emph{et~al.}(1994)Siag, Tennal, and Mazumder]{Siag1994}
A.~Siag, K.~Tennal and M.~Mazumder, \emph{Particul. Sci. Technol.}, 1994,
  \textbf{12}, 351--365\relax
\mciteBstWouldAddEndPuncttrue
\mciteSetBstMidEndSepPunct{\mcitedefaultmidpunct}
{\mcitedefaultendpunct}{\mcitedefaultseppunct}\relax
\EndOfBibitem
\bibitem[Yim \emph{et~al.}(2020)Yim, Cheng, Patel, Kui, Meng, and
  Jokerst]{Yim2020}
W.~Yim, D.~Cheng, S.~Patel, R.~Kui, Y.~S. Meng and J.~V. Jokerst, \emph{medRxiv
  preprint 2020.07.07.20148551}, 2020\relax
\mciteBstWouldAddEndPuncttrue
\mciteSetBstMidEndSepPunct{\mcitedefaultmidpunct}
{\mcitedefaultendpunct}{\mcitedefaultseppunct}\relax
\EndOfBibitem
\bibitem[Effros \emph{et~al.}(2002)Effros, Hoagland, Bosbous, Castillo, Foss,
  Dunning, Gare, Lin, and Sun]{Effros2002}
R.~M. Effros, K.~W. Hoagland, M.~Bosbous, D.~Castillo, B.~Foss, M.~Dunning,
  M.~Gare, W.~Lin and F.~Sun, \emph{Am. J. Respir. Crit. Care Med.}, 2002,
  \textbf{165}, 663--669\relax
\mciteBstWouldAddEndPuncttrue
\mciteSetBstMidEndSepPunct{\mcitedefaultmidpunct}
{\mcitedefaultendpunct}{\mcitedefaultseppunct}\relax
\EndOfBibitem
\bibitem[Vejerano and Marr(2018)]{Vejerano2018}
E.~P. Vejerano and L.~C. Marr, \emph{J. R. Soc. Interface}, 2018, \textbf{15},
  20170939\relax
\mciteBstWouldAddEndPuncttrue
\mciteSetBstMidEndSepPunct{\mcitedefaultmidpunct}
{\mcitedefaultendpunct}{\mcitedefaultseppunct}\relax
\EndOfBibitem
\bibitem[Markovich \emph{et~al.}(2020)Markovich, Andelman, and
  Podgornik]{Safinya}
T.~Markovich, D.~Andelman and R.~Podgornik, in \emph{Handbook of Lipid
  Membranes}, ed. C.~Safinya and J.~Raedler, Taylor \& Francis, 2020,
  ch.~9\relax
\mciteBstWouldAddEndPuncttrue
\mciteSetBstMidEndSepPunct{\mcitedefaultmidpunct}
{\mcitedefaultendpunct}{\mcitedefaultseppunct}\relax
\EndOfBibitem
\bibitem[Lu \emph{et~al.}(2005)Lu, Cheng, Hou, and McCammon]{Lu_JCP_2005}
B.~Lu, X.~Cheng, T.~Hou and J.~A. McCammon, \emph{J. Chem. Phys.}, 2005,
  \textbf{123}, 084904\relax
\mciteBstWouldAddEndPuncttrue
\mciteSetBstMidEndSepPunct{\mcitedefaultmidpunct}
{\mcitedefaultendpunct}{\mcitedefaultseppunct}\relax
\EndOfBibitem
\bibitem[Gilson \emph{et~al.}(1993)Gilson, Davis, Luty, and
  McCammon]{Gilson_JPC_1993}
M.~K. Gilson, M.~E. Davis, B.~A. Luty and J.~A. McCammon, \emph{J. Phys.
  Chem.}, 1993, \textbf{97}, 3591--3600\relax
\mciteBstWouldAddEndPuncttrue
\mciteSetBstMidEndSepPunct{\mcitedefaultmidpunct}
{\mcitedefaultendpunct}{\mcitedefaultseppunct}\relax
\EndOfBibitem
\bibitem[Johnson \emph{et~al.}(1973)Johnson, Wagner, and Bancroft]{Johnson1973}
M.~Johnson, G.~Wagner and J.~Bancroft, \emph{J. Gen. Virol.}, 1973,
  \textbf{19}, 263--273\relax
\mciteBstWouldAddEndPuncttrue
\mciteSetBstMidEndSepPunct{\mcitedefaultmidpunct}
{\mcitedefaultendpunct}{\mcitedefaultseppunct}\relax
\EndOfBibitem
\bibitem[Heffron and Mayer(2021)]{Heffron2021}
J.~Heffron and B.~K. Mayer, \emph{Appl. Environ. Microbiol.}, 2021,
  \textbf{87}, e02319--20\relax
\mciteBstWouldAddEndPuncttrue
\mciteSetBstMidEndSepPunct{\mcitedefaultmidpunct}
{\mcitedefaultendpunct}{\mcitedefaultseppunct}\relax
\EndOfBibitem
\bibitem[Cats \emph{et~al.}(2020)Cats, Evans, Härtel, and van Roij]{cats2021}
P.~Cats, R.~Evans, A.~Härtel and R.~van Roij, \emph{arXiv preprint 2012.02713
  [cond-mat.soft]}, 2020\relax
\mciteBstWouldAddEndPuncttrue
\mciteSetBstMidEndSepPunct{\mcitedefaultmidpunct}
{\mcitedefaultendpunct}{\mcitedefaultseppunct}\relax
\EndOfBibitem
\bibitem[Roth \emph{et~al.}(1996)Roth, Neal, and Lenhoff]{Lenhoff1996}
C.~Roth, B.~Neal and A.~Lenhoff, \emph{Biophys. J.}, 1996, \textbf{70},
  977--987\relax
\mciteBstWouldAddEndPuncttrue
\mciteSetBstMidEndSepPunct{\mcitedefaultmidpunct}
{\mcitedefaultendpunct}{\mcitedefaultseppunct}\relax
\EndOfBibitem
\bibitem[Podgornik and Andelman(2020)]{Embarras2020}
R.~Podgornik and D.~Andelman, \emph{Embarras de richesses in non-DLVO colloid
  interactions}, 2020,
  \url{http://dx.doi.org/10.36471/JCCM_September_2020_02}\relax
\mciteBstWouldAddEndPuncttrue
\mciteSetBstMidEndSepPunct{\mcitedefaultmidpunct}
{\mcitedefaultendpunct}{\mcitedefaultseppunct}\relax
\EndOfBibitem
\bibitem[Hernando-P{\'e}rez \emph{et~al.}(2015)Hernando-P{\'e}rez,
  Cartagena-Rivera, Bo{\v{z}}i{\v{c}}, Carrillo, San~Mart{\'\i}n, Mateu, Raman,
  Podgornik, and De~Pablo]{Hernando2015}
M.~Hernando-P{\'e}rez, A.~Cartagena-Rivera, A.~Bo{\v{z}}i{\v{c}}, P.~J.
  Carrillo, C.~San~Mart{\'\i}n, M.~G. Mateu, A.~Raman, R.~Podgornik and
  P.~De~Pablo, \emph{Nanoscale}, 2015, \textbf{7}, 17289--17298\relax
\mciteBstWouldAddEndPuncttrue
\mciteSetBstMidEndSepPunct{\mcitedefaultmidpunct}
{\mcitedefaultendpunct}{\mcitedefaultseppunct}\relax
\EndOfBibitem
\end{mcitethebibliography}


\begin{thebibliography}{5}%
\makeatletter
\providecommand \@ifxundefined [1]{%
 \@ifx{#1\undefined}
}%
\providecommand \@ifnum [1]{%
 \ifnum #1\expandafter \@firstoftwo
 \else \expandafter \@secondoftwo
 \fi
}%
\providecommand \@ifx [1]{%
 \ifx #1\expandafter \@firstoftwo
 \else \expandafter \@secondoftwo
 \fi
}%
\providecommand \natexlab [1]{#1}%
\providecommand \enquote  [1]{``#1''}%
\providecommand \bibnamefont  [1]{#1}%
\providecommand \bibfnamefont [1]{#1}%
\providecommand \citenamefont [1]{#1}%
\providecommand \href@noop [0]{\@secondoftwo}%
\providecommand \href [0]{\begingroup \@sanitize@url \@href}%
\providecommand \@href[1]{\@@startlink{#1}\@@href}%
\providecommand \@@href[1]{\endgroup#1\@@endlink}%
\providecommand \@sanitize@url [0]{\catcode `\\12\catcode `\$12\catcode
  `\&12\catcode `\#12\catcode `\^12\catcode `\_12\catcode `\%12\relax}%
\providecommand \@@startlink[1]{}%
\providecommand \@@endlink[0]{}%
\providecommand \url  [0]{\begingroup\@sanitize@url \@url }%
\providecommand \@url [1]{\endgroup\@href {#1}{\urlprefix }}%
\providecommand \urlprefix  [0]{URL }%
\providecommand \Eprint [0]{\href }%
\providecommand \doibase [0]{http://dx.doi.org/}%
\providecommand \selectlanguage [0]{\@gobble}%
\providecommand \bibinfo  [0]{\@secondoftwo}%
\providecommand \bibfield  [0]{\@secondoftwo}%
\providecommand \translation [1]{[#1]}%
\providecommand \BibitemOpen [0]{}%
\providecommand \bibitemStop [0]{}%
\providecommand \bibitemNoStop [0]{.\EOS\space}%
\providecommand \EOS [0]{\spacefactor3000\relax}%
\providecommand \BibitemShut  [1]{\csname bibitem#1\endcsname}%
\let\auto@bib@innerbib\@empty
\bibitem [{\citenamefont {Walls}\ \emph {et~al.}(2020)\citenamefont {Walls},
  \citenamefont {Park}, \citenamefont {Tortorici}, \citenamefont {Wall},
  \citenamefont {Mcguire},\ and\ \citenamefont {Veesler}}]{Walls2020}%
  \BibitemOpen
  \bibfield  {author} {\bibinfo {author} {\bibfnamefont {A.}~\bibnamefont
  {Walls}}, \bibinfo {author} {\bibfnamefont {Y.-J.}\ \bibnamefont {Park}},
  \bibinfo {author} {\bibfnamefont {M.~A.}\ \bibnamefont {Tortorici}}, \bibinfo
  {author} {\bibfnamefont {A.}~\bibnamefont {Wall}}, \bibinfo {author}
  {\bibfnamefont {A.}~\bibnamefont {Mcguire}}, \ and\ \bibinfo {author}
  {\bibfnamefont {D.}~\bibnamefont {Veesler}},\ }\href@noop {} {\bibfield
  {journal} {\bibinfo  {journal} {Cell}\ }\textbf {\bibinfo {volume} {181}},\
  \bibinfo {pages} {281} (\bibinfo {year} {2020})}\BibitemShut {NoStop}%
\bibitem [{\citenamefont {Olsson}\ \emph {et~al.}(2011)\citenamefont {Olsson},
  \citenamefont {S{\o}ndergaard}, \citenamefont {Rostkowski},\ and\
  \citenamefont {Jensen}}]{propka3}%
  \BibitemOpen
  \bibfield  {author} {\bibinfo {author} {\bibfnamefont {M.~H.}\ \bibnamefont
  {Olsson}}, \bibinfo {author} {\bibfnamefont {C.~R.}\ \bibnamefont
  {S{\o}ndergaard}}, \bibinfo {author} {\bibfnamefont {M.}~\bibnamefont
  {Rostkowski}}, \ and\ \bibinfo {author} {\bibfnamefont {J.~H.}\ \bibnamefont
  {Jensen}},\ }\href@noop {} {\bibfield  {journal} {\bibinfo  {journal} {J.
  Chem. Theory Comput.}\ }\textbf {\bibinfo {volume} {7}},\ \bibinfo {pages}
  {525} (\bibinfo {year} {2011})}\BibitemShut {NoStop}%
\bibitem [{\citenamefont {Bo\v{z}i\v{c}}\ and\ \citenamefont
  {Podgornik}(2017)}]{Bozic2017}%
  \BibitemOpen
  \bibfield  {author} {\bibinfo {author} {\bibfnamefont {A.}~\bibnamefont
  {Bo\v{z}i\v{c}}}\ and\ \bibinfo {author} {\bibfnamefont {R.}~\bibnamefont
  {Podgornik}},\ }\href@noop {} {\bibfield  {journal} {\bibinfo  {journal}
  {Biophys. J.}\ }\textbf {\bibinfo {volume} {113}},\ \bibinfo {pages} {1454}
  (\bibinfo {year} {2017})}\BibitemShut {NoStop}%
\bibitem [{\citenamefont {Lide}(2013)}]{CRC-94}%
  \BibitemOpen
  \bibinfo {editor} {\bibfnamefont {D.~R.}\ \bibnamefont {Lide}},\ ed.,\
  \href@noop {} {\emph {\bibinfo {title} {CRC handbook of Chemistry and
  Physics}}},\ \bibinfo {edition} {94th}\ ed.\ (\bibinfo  {publisher} {CRC
  Press},\ \bibinfo {address} {Boston},\ \bibinfo {year} {2013})\BibitemShut
  {NoStop}%
\bibitem [{\citenamefont {Markovich}\ \emph {et~al.}(2020)\citenamefont
  {Markovich}, \citenamefont {Andelman},\ and\ \citenamefont
  {Podgornik}}]{Safinya}%
  \BibitemOpen
  \bibfield  {author} {\bibinfo {author} {\bibfnamefont {T.}~\bibnamefont
  {Markovich}}, \bibinfo {author} {\bibfnamefont {D.}~\bibnamefont {Andelman}},
  \ and\ \bibinfo {author} {\bibfnamefont {R.}~\bibnamefont {Podgornik}},\
  }\enquote {\bibinfo {title} {Handbook of lipid membranes},}\ \ (\bibinfo
  {publisher} {Taylor \& Francis},\ \bibinfo {year} {2020})\ Chap.\ \bibinfo
  {chapter} {Charged Membranes: Poisson-Boltzmann theory, DLVO paradigm and
  beyond}\BibitemShut {NoStop}%
\end{thebibliography}%
\bibliographystyle{rsc} 

\end{document}


\title{Electronic Supplementary Information:\\Electrostatic interaction between SARS-CoV-2 virus and charged electret fibre}

\author{Leili Javidpour}
\affiliation{School of Physics, Institute for Research in Fundamental Sciences (IPM), Tehran, 19395-5531, Iran}
\author{An\v ze Bo\v zi\v c}
\affiliation{Department of Theoretical Physics, Jo\v{z}ef Stefan Institute, SI-1000 Ljubljana, Slovenia}
\author{Ali Naji}
\affiliation{School of Physics, Institute for Research in Fundamental Sciences (IPM), Tehran, 19395-5531, Iran}
\affiliation{School of Nano Science, Institute for Research in Fundamental Sciences (IPM), Tehran, 19395-5531, Iran}
\author{Rudolf Podgornik}
\email{podgornikrudolf@ucas.ac.cn}
\thanks{Also affiliated with Kavli Institute for Theoretical Sciences, University of Chinese Academy of Sciences, Beijing 100049, China,  Department of Physics, Faculty of Mathematics and Physics, University of Ljubljana, SI-1000 Ljubljana, Slovenia and Department of Theoretical Physics, Jo\v zef Stefan Institute, SI-1000 Ljubljana, Slovenia}
\affiliation{School of Physical Sciences, University of Chinese Academy of Sciences, Beijing 100049, China and CAS Key Laboratory of Soft Matter Physics, Institute of Physics, Chinese Academy of Sciences, Beijing 100190, China}
\affiliation{Wenzhou Institute of the University of Chinese Academy of Sciences, Wenzhou, Zhejiang 325000, China}


\maketitle

\renewcommand\thefigure{S\arabic{figure}}
\setcounter{figure}{0}
\renewcommand\thetable{S\Roman{table}}   
\setcounter{table}{0}

\section{Charge regulation parameters}

To determine the charge on the S proteins, we first extract solvent-accessible ionizable amino acids (AA) from the structural data in PDB:6VXX~\cite{Walls2020}. We consider deprotonated (ASP, GLU, TYR, and CYS) and protonated (ARG, LYS, and HIS) AAs, whose solvent accessibility is determined using propKa 3.1~\cite{propka3}. Only those AA residues that are buried less than $80\%$ are kept and considered to carry any charge~\cite{Bozic2017}. This retains $504$ out of $702$ ionizable AA residues, and the detailed composition by residue type is given in Table~\ref{table:tab1}. Given the general coarse-grained nature of our model, we attribute to each AA functional group its $\pK$ value in bulk dilute aqueous solution. The values in Table~\ref{table:tab1} are then used in Eq.~(2) in the main text together with any local electrostatic potential to obtain the charge-regulated surface charge distribution on the model S proteins.

\begin{table}[!htp]
\begin{center}
\caption{Charge regulation parameters for the surface of S proteins. For each deprotonated ($q_\pm =-1$ $e$) and protonated ($q_\pm=+1$ $e$) AA functional group we use their $\pK$ value in bulk dilute aqueous solution, taken from Ref.~\cite{CRC-94}. The number of ionizable, solvent-accessible amino acid residues $N_S$ on the S protein is obtained from structural data PDB:6VXX.}
\begin{tabular}{c|c|c|c} 
AA & $q_\pm$ [$e$] & pK$_\mathrm{a}$ & $N_{S}$  \\
\hline
\hline
ASP & $-1$ & $3.71$ & $123$ \\
GLU & $-1$ & $4.15$ & $84$ \\
TYR & $-1$ & $10.10$ & $84$ \\
CYS & $-1$ & $8.14$ & $3$ \\
ARG & $+1$ & $12.10$ & $63$ \\
LYS & $+1$ & $10.67$ & $120$ \\
HIS & $+1$ & $6.04$ & $27$ \\
\end{tabular}
\label{table:tab1}
\end{center}
\end{table}

\section{Electrostatic potential}

\begin{figure}[!htp]
\begin{center}
\includegraphics[width=0.45\textwidth]{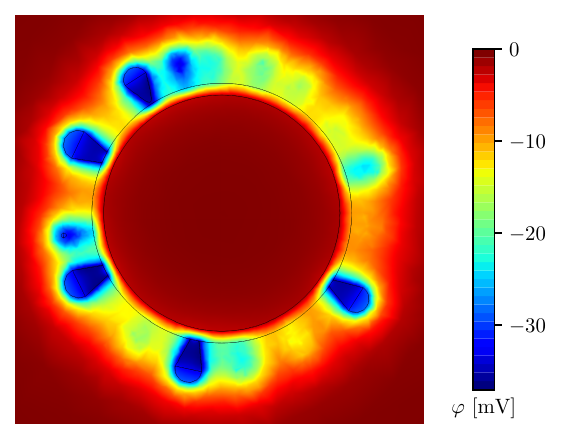}
\caption{Two-dimensional cross-section of the model, showing the local electrostatic potential $\varphi$ of a virus with $N=80$ S proteins in the absence of the electret surface, with $n_0=2$ mM and $\pH=7$. Black circles show the inner and outer surface cross-sections of the lipid membrane, which carry no charge.
}
\label{fig:S2}
\end{center}
\end{figure}

The local electrostatic potential is obtained as a solution of the PB equation [Eq.~(3) in the main text] in the regions accessible to the bathing solution (external solution accommodating the virus as well as the virus interior) and the Poisson equation in the inaccessible regions (membrane envelope, S protein cores, and the electret substrate), with boundary conditions given by Eq.~(4) in the main text. Since the surface charge density on the S proteins is obtained from the charge regulation model, it varies spatially along the surface; on the other hand, the surface charge density on the electret is assumed to be fixed. Example in Fig.~\ref{fig:S2} shows a two-dimensional cross-section of the local electrostatic potential around the virus in the absence of the electret substrate, i.e., for an isolated virus.

\section{Maxwell and osmotic components of the total electrostatic force}

Based on the standard PB theory~\cite{Safinya}, the stress tensor anywhere within the bathing ion solution can be decomposed into Maxwell and van't Hoff components, corresponding to the field stresses and the salt ion osmotic pressure, respectively [Eq.~(5) in the main text]. The total force on a body within the bathing solution is then obtained by integrating the stress tensor over any closed surface containing the body. Numerically, it is best to choose an integration surface that is finely meshed to ensure a high numerical accuracy of the result, and to obtain our results, we have consistently chosen the surface next to the epitope of the virus.

\begin{figure}[!htp]
\begin{center}
\includegraphics[width=0.65\textwidth]{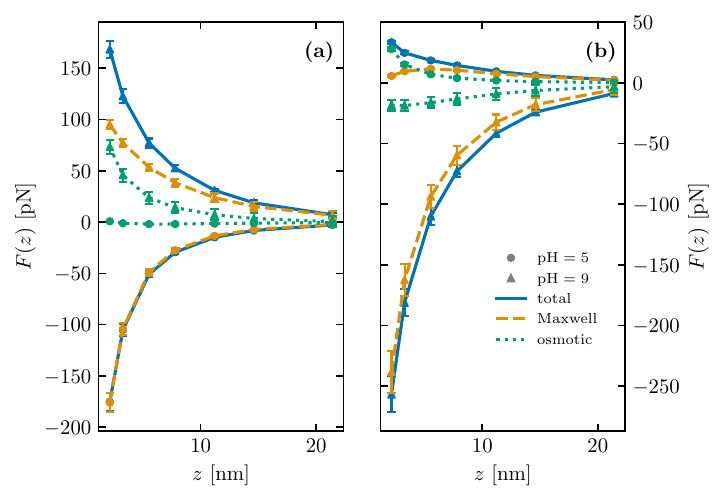}
\caption{Total electrostatic force on the virus and its Maxwell (electrostatic) and osmotic components as a function of distance from the electret surface, for a system with $N=80$, $n_0=2$ mM, and {\bf (a)} $\sigma_S=-1$ $e/\mathrm{nm}^2$ and {\bf (b)} $\sigma_S=+1$ $e/\mathrm{nm}^2$. The legend is the same for both panels. Shown are also the errorbars, obtained by averaging over approximately $20$ different Mitchell configurations of S proteins on the surface of the lipid membrane for each data point.
}
\label{fig:S3}
\end{center}
\end{figure}

The relative contributions of Maxwell and van't Hoff components to the total stress tensor---and consequently to the total interaction force---vary with the external parameters. Figure~\ref{fig:S3} shows the total electrostatic force on the virus as a function of its distance from the electret surface together with its decomposition into the electrostatic component from the Maxwell stress tensor and the osmotic component. When the net force is attractive, the Maxwell component clearly dominates, whereas when the net force is repulsive, the contributions of the Maxwell and van't Hoff terms are comparable, regardless of the sign of the surface charge density on the electret fibre.

\bibliography{references}